\documentstyle[epsfig,12pt]{article}
\newcommand{\be}{\begin{eqnarray}}
\newcommand{\ee}{\end{eqnarray}}
\newcommand{\ba}{\begin{array}}
\newcommand{\ea}{\end{array}}
\newcommand{\ur}[1]{(\ref{#1})}

\newcommand{\eq}[1]{eq.~(\ref{#1})}
\newcommand{\eqs}[2]{eqs.(\ref{#1}, \ref{#2})}
\newcommand{\Eq}[1]{Eq.~(\ref{#1})}

\newcommand{\eqss}[3]{eqs.(\ref{#1}, \ref{#2}, \ref{#3})}

\newcommand{\Tr}{{\rm Tr}\,}

\newcommand{\laa}{\langle\!\langle}
\newcommand{\raa}{\rangle\!\rangle}

 \def\Sp{\mbox{Sp}}

  \newcommand{\la}[1]{\label{#1}}
  
  \def\beq{\begin{equation}}
  \def\eeq{\end{equation}}
  \def\beqr{\begin{eqnarray}}
  \def\eeqr{\end{eqnarray}}

\begin{document}
\thispagestyle{empty}
%\today
\vskip 3.5true  cm
\begin{center}
{\Large\bf Trying to Understand Confinement \\
\vskip .5true cm
in the Schr\"odinger Picture
\footnote{Lecture at the 4-th St. Petersburg Winter School in
Theoretical Physics, Feb. 22-28, 1998.}
} \\
\vskip 1.5true cm

{\large\bf Dmitri Diakonov}\\
\vskip 1true cm
{\it NORDITA, Blegdamsvej 17, 2100 Copenhagen \O, Denmark}\\
\vskip .5true cm
and\\
\vskip .5true cm
{\it Petersburg Nuclear Physics Institute, Gatchina,
St.Petersburg 188350, Russia \\}
\end{center}
\vskip 1.5true cm
\begin{abstract}
\noindent We study the gauge-invariant gaussian ansatz for the vacuum
wave functional and show that it potentially posseses many desirable
features of the Yang--Mills theory, like asymptotic freedom,
mass generation through the transmutation of dimensions and a linear
potential between static quarks. We point out that these (and
other) features can be studied in a systematic way by combining
perturbative and $1/n$ expansions. Contrary to the euclidean approach,
confinement can be easily formulated and easily built in, if not derived,
in the variational Schr\"odinger approach.
\end{abstract}
\newpage

\section{Introduction}

\noindent We shall consider the pure Yang--Mills theory, else called
gluodynamics, with no dynamical quarks. It is widely believed
that in this theory the potential between probe static quarks
is infinitely rising linearly with the quark separation. If proved
correct, it will be one of the most remarkable findings in physics,
and one would need to explain it from different angles. The present
paper is an attempt to understand the physics of confinement
in the framework of the Schr\"odinger picture.

The use of the Schr\"odinger equation to discuss the ground state of
gluodynamics as well as the two source quarks on top of the ground
state, is not new. To mention a few early works, see
refs.\cite{Greensite,Fey,Corn,Mansf}. More recently, the interest to
the Schr\"odinger equation has been revived by Kogan and Kovner who
suggested a variational vacuum wave functional in a gaussian form
averaged over all gauge transformations \cite{KoKo1}. We shall use this
form of the trial wave functional making a slight but essential
generalization of the ansatz suggested in \cite{KoKo1}. This work has
been influenced also by a recent paper by Zarembo \cite{Zar1} though we
differ from that reference in several aspects.

\section{Schr\"odinger equation}

The spatial components of the Yang--Mills potential $A_i^a(x)$
play the role of coordinates, the electric field strength $E_i^a(x)
= \dot A_i^a(x)$ plays the role of momenta. The classical energy of
the Yang--Mills field is

\beq
{\cal H}^{class}
=\int \!d^3x\left[\frac{1}{2g^2}(E_i^a(x))^2
+\frac{1}{2g^2}(B_i^a(x))^2\right]
\la{classen}\eeq
where

\beq
B_i^a=\epsilon_{ijk}(\partial_jA_k^a+(1/2)f^{abc}A_j^bA_k^c)
\la{B}\eeq
is the magnetic field strength. Upon quantization the
electric field is replaced by variational derivative, $E_i^a(x)
\rightarrow -ig^2\delta/\delta A_i^a(x)$, if one uses the `coordinate
representation' for the wave functional. The functional Schr\"odinger
equation for the wave functional $\Psi[A_i^a(x)]$ takes the form

\beq
{\cal H}\Psi[A_i]=
\int \!d^3x\left\{-\frac{g^2}{2}\frac{\delta^2}{(\delta A_i^a(x))^2}
+\frac{1}{2g^2}(B_i^a(x))^2\right\}\Psi[A_i]
={\cal E}\Psi[A_i]
\la{schr1}\eeq
where ${\cal E}$ is the eigenenergy of the state in question.

The Schr\"odinger equation \ur{schr1} should be supplemented by the
gauge condition. For a state without external colour sources the
condition is that the wave functional $\Psi[A_i]$ should be
gauge-invariant,

\beq
\Psi[A_i^S]=\Psi[A_i],
\la{gauge-inv}\eeq
where

\beq
A_i^S\equiv S^\dagger A_iS+iS^\dagger\partial_iS
\la{gauge-trans}\eeq
is the gauge transformation. For a state with an external static quark
at $z_1$ and an antiquark at  $z_2$ the wave functional
$\Psi^\alpha_\beta[A_i;z_1,z_2]$ should satisfy the gauge condition

\beq
\Psi^\alpha_\beta[A_i^S;z_1,z_2]
=\left[S^\dagger(z_1)\right]^\alpha_{\alpha^\prime}
\Psi^{\alpha^\prime}_{\beta^\prime}[A_i;z_1,z_2]
\left[S(z_2)\right]^{\beta^\prime}_\beta .
\la{gauge-cov}\eeq
If the probe sources belong to another representation the
gauge-tranforming matrices $S(z_{1,2})$ should be taken in appropriate
representation.

\section{Ultraviolet regularization}

Strictly speaking, the hamiltonian \ur{schr1} is senseless since
any calculations with it are plagued by ultraviolet divergences.
In order to give meaning to the Schr\"odinger equation \ur{schr1}
one has to regularize both the kinetic and potential energy parts
of the hamiltonian. Examples of ultraviolet regularization are provided
by the lattice hamiltonian \cite{KogSus} or by the `heat
kernel' method \cite{Mansf}.

The requirements on possible ultraviolet regularizations are rather
stringent. In addition to the apparent requirement that the
regularization should be gauge invariant, there is a less trivial
requirement that the regularized hamiltonian, together with the
momentum and angular momentum operators, should satisfy the Poincar\`e
algebra. Otherwise, there would be various inconsistencies, like
that the energy of a glueball state would not obey the relativistic
formula $E^2=m^2+p^2$. To my knowledge, a consistent regularization
scheme has not been yet introduced.

Our aim is, however, the vacuum state with and without static sources,
therefore any of the two mentioned regularization schemes would suit
us here. In fact, one has to solve the regularized Schr\"odinger
equation

\beq
{\cal H}^{reg}(g,M)\Psi(g,M)={\cal E}(g,M)\Psi(g,M)
\la{schr2}\eeq
where $g$ is the bare coupling constant given at certain large
normalization scale $M$ which one should eventually put to infinity.
[For example, $M$ can be viewed as the inverse lattice spacing in
case of the lattice regularization].

Let us consider the ground state or vacuum of the theory. Its energy
${\cal E}$ is by far dominated by the energy of the normal zero-point
oscillations of the gauge field, which diverges quartically and, being
regularized, is proportional to $M^4$. From the perturbation theory
one infers

\[
{\cal E}^{pert}(g,M)=2(N_c^2-1)\sum
\left(\frac{\hbar\omega}{2}\right)
+\begin{array}{c}{\rm perturbative} \\ {\rm corrections}
\end{array}
\]
\[
=2(N_c^2-1)V\int\!\frac{d^3k}{(2\pi)^3}\frac{|k|}{2}
+\begin{array}{c}{\rm perturbative} \\ {\rm corrections}
\end{array}
\]
\beq
=VM^4(c_{00}+c_{01}g^2+c_{02}g^4+...).
\la{Epert}\eeq
It is not the full truth, however. We expect also {\em
non-perturbative} contributions to the vacuum energy determined
by the dimensional parameter $\Lambda$ obtained through the
transmutation of dimensions,

\beq
\Lambda=M\left(\frac{bg^2}{16\pi^2}\right)^{-\frac{b^\prime}{2b^2}}
\exp\left(-\frac{8\pi^2}{bg^2}\right)\left(1+O(g^2)\right),\;\;\;\;\;
b=\frac{11}{3}N_c,\;\;\;\;\;b^\prime=\frac{34}{3}N_c^2,
\la{Lambda}\eeq
where $b$ is the one-loop and $b^\prime$ is the two-loop Gell-Mann--Low
coefficient. Contributions proportional to powers of $\Lambda$ cannot
be obtained in any order of the perturbation theory. However,
there are no reasons why such contributions should be absent.
Therefore, we write for the vacuum energy a general double expansion:

\[
\frac{{\cal E}(g,M)}{V}=M^4(c_{00}+c_{01}g^2+c_{02}g^4+...)
+M^3\Lambda(c_{10}+c_{11}g^2+...)+M^2\Lambda^2c_{20}+...
+M\Lambda^3c_{30}+...
\]
\beq
+\Lambda^4(c_{40}+c_{41}g^2+...).
\la{Eexpansion}\eeq
Meanwhile, all physical observables, like the string tension in the
static potential, glueball masses, gluon condensate, topological
susceptibility, etc., should arise as renorm-invariant combinations of
the bare coupling $g$ and the ultraviolet cutoff $M$, {\em i.e.} should
be expressed trough $\Lambda$ only, in appropriate powers. This makes a
straightforward use of the variational principle to extract physical
observables almost a hopeless task: before getting to physically
interesting quantities encoded in the last term in \eq{Eexpansion}
one has to minimize the first four terms which are by far the larger!

There is an interesting possibility to circumvent this difficulty,
at least for the vacuum state. Let us assume that we have succeeded in
building the regularized operators for the full stress-energy tensor,
$\theta_{\mu\nu}=[F^a_{\mu\alpha}F^a_{\nu\alpha}
-(1/4)g_{\mu\nu}F^a_{\gamma\alpha}F^a_{\gamma\alpha}]/g^2$,
in accordance with the Poincar\`e algebra, the hamiltonian being
${\cal H}=\int\! d^3x\:\theta_{00} = \int\!d^3x\:(E^2+B^2)/2g^2$.
The vacuum energy is ${\cal E}=V\langle \theta_{00}\rangle$ where
$\langle ...\rangle$ denotes averaging over the vacuum wave functional.

Because of the Lorentz invariance of the vacuum one expects
$\langle \theta_{00}\rangle=-\langle \theta_{11}\rangle=
-\langle \theta_{22}\rangle = -\langle \theta_{33}\rangle$ and hence
$\langle \theta_{00}\rangle=(1/4)\langle \theta_{\mu\mu}\rangle$.
The last quantity is naively zero (following from the above definition
of $\theta_{\mu\nu}$), however under proper regularization it
should be equal, through trace anomaly, to

\beq
\theta_{\mu\mu}=\frac{\beta(g^2)}{4g^4}\frac{(F^a_{\mu\nu})^2}{32\pi^2}
=-b\frac{B^2-E^2}{16\pi^2}.
\la{sled}\eeq
[We have neglected higher-loop contributions to the Gell-Mann--Low
function $\beta(g^2)$ which can always be done if the ultraviolet
cutoff $M$ is chosen large enough and hence the bare coupling
$g$ small enough.] The quantity
$\langle (F^a_{\mu\nu})^2\rangle$ is known as the gluon
condensate appearing in the ITEP sum rules and has to be
renorm-invariant and hence proportional to $\Lambda^4$.

Thus if one wants to extract directly the non-perturbative physics
from the variational approach one should rather maximize the gluon
condensate $\langle B^2-E^2\rangle$ than minimize the energy
density $(B^2+E^2)/2g^2$ \footnote{This is what one actualy does
in building the euclidean instanton vacuum from the variational
principle \cite{DP}, where one is also interested in finding the
shift of the vacuum energy in respect to the perturbative one.}.

In the latter quantity the kinetic and potential energies of the
zero-point gluon oscillations add up and diverge quartically, see
\eq{Epert}; in the former quantity they cancel each other. In the zero
order of the perturbation theory this cancellation is exact due to the
virial theorem for harmonic oscillators. Under a proper regularization
of the operators, respecting the $SO(4,2)$ generalization of the
Poincar\`e algebra (including the dilatation and conformal operators),
this cancellation should hold to any order of the perturbation theory,
so that the properly defined gluon condensate gets a non-zero
contribution only from the non-perturbative physics which is
characterized by being non-analytic in the coupling constant, see
\eq{Lambda}.

Unfortunately, the prescription to maximize $\langle B^2-E^2\rangle$
is {\it a priori} not well-defined either as one can always invent a
crazy trial functional making the quantity as large as one likes.

Despite these pessimistic remarks concerning the foundations for
the use of the variational principle,
it may still serve as a useful guide to understand non-perturbative
physics in a language different from the usual euclidean approach.
Being unable to formulate the problem in a mathematically unequivocal
way, we shall henceforth proceed in a somewhat sloppy manner: we shall
not regularize the hamiltonian but simply cut the divergent momenta
integrals `by hands' at certain large momentum $M\gg\Lambda$. We shall
not go beyond the one-loop calculations, so this barbarous procedure
will be sufficient for our needs. In fact, we shall be able to separate
the perturbative and nonperturbative physics using a specific
variational ansatz, and minimizing the nonperturbative vacuum energy
will make sense, see section 8.

\section{Kogan--Kovner variational ansatz and its generalization}

Kogan and Kovner \cite{KoKo1} have suggested a gaussian trial wave
functional for the vacuum state, averaged over all gauge
transformations of the `coordinates' $A_i$ :

\beq
\Psi[A]=\int \!DU(x)\exp\left(-\frac{1}{g^2}Q[A^U]\right)
\la{KoKo}\eeq
where $A^U_i\equiv U^\dagger A_i U+iU^\dagger \partial_i U$ is the
gauge-transformed Yang--Mills potential and $Q[A]$ is a
{\em quadratic} functional (hence the notation):

\beq
Q[A]=\int\!\int\!d^3x\:d^3y\:K_{ij}(x-y)\:\Tr A_i(x)A_j(y).
\la{Q}\eeq
The kernel $K_{ij}(x-y)$ is the only variational (or trial) function of
the ansatz. Its Fourier transform has the general structure compatible
with space isotropy:

\beq
K_{ij}(p)=a(|p|)\cdot\delta_{ij}+b(|p|)\cdot\frac{p_ip_j}{p^2}.
\la{Kp}\eeq

The two scalar functions $a(p)$ and $b(p)$ are the
variational `parameters' of the ansatz. As we shall see below,
minimization of the vacuum energy in the leading order of the
perturbation theory (the coefficient $c_{00}$ in the notations
of \eq{Eexpansion}) leads to the following result:

\beq
a^{pert}(p)=|p|,\;\;\;\;\;\; b^{pert}(p)={\rm arbitrary}.
\la{abpert}\eeq

In ref. \cite{KoKo1} the longitudinal part of the kernel $b(p)$ has been
put to zero (without compelling reasons). In ref. \cite{Zar1} it has
been shown that, because of the averaging over gauges in the wave
functional \ur{KoKo}, certain forms of ``actions'' in the exponent of
$\Psi[A]$ are physically equivalent; in particular, the longitudinal
part of the quadratic form $Q[A]$ can be removed by choosing another
``action'' in the exponent, containing higher powers of $A_i$. Having
no objections to this statement in general we note, however, that if
one resctricts oneself to purely quadratic ``actions'' of the form
\ur{Q} the longitudinal part of $Q[A_i]$ becomes physically meaningful.
Moreover, we shall show below that by suitably choosing the
longitudinal function $b(p)$ in \ur{Kp} one can reproduce the correct
$(11/3)N_c$ coefficient in the one-loop Gell-Mann--Low function. In
ref. \cite{BrownKo} the value $(12/3)N_c$ has been obtained instead for
that coefficient, with $b(p)$ set to zero.

According to the variational principle the vacuum energy satisfies
the inequality

\beq
{\cal E}\geq \frac{\langle \Psi|{\cal H}|\Psi\rangle}
{\langle \Psi|\Psi\rangle}.
\la{meanenergy}\eeq
One has to minimize the r.h.s. in the free parameters / functions
of the chosen ansatz, in our case, in the two scalar functions $a(p)$
and $b(p)$.

\section{Evaluation of the norm ${\cal N}=\langle \Psi|\Psi\rangle $}

The norm ${\cal N}$ standing in the denominator of \eq{meanenergy}
is understood as the functional integral,

\[
{\cal N}=\langle \Psi|\Psi\rangle=\int\!DA\;\Psi^*[A]\Psi[A]
\]
\beq
=\int\!DA\int\!DU_1\int\!DU_2\exp\left(-\frac{1}{g^2}Q[A^{U_1}_i]
-\frac{1}{g^2}Q[A^{U_2}_i]\right).
\la{norm1}\eeq

One first integrates over the gauge potential $A$ by saddle-point
method; since $Q[A^{U_{1,2}}_i]$ are quadratic in $A$ the
saddle-point integration is exact.

We have

\beq
\frac{\delta Q[A^U]}{\delta A_i^a(x)}=
\int\!d^3y\: K_{ij}(x-y)\:2\:\Tr U^\dagger(x)t^aU(x)
\left[U^\dagger(y)A_j(y)U(y)+iU^\dagger(y)\partial_jU(y)\right]
\la{deltaQ}\eeq
where $t^a$ are $SU(N_c)$ generators in the fundamental
representation normalized to $\Tr t^at^b =(1/2)\delta_{ab}$.
The saddle-point equation for $A_i$ is

\beq
\frac{\delta Q[A^{U_1}]}{\delta A_i^a(x)}
+\frac{\delta Q[A^{U_2}]}{\delta A_i^a(x)}=0.
\la{sp0}\eeq
The solution of this equation gives the saddle-point value
of $A_i$:

\beq
\bar A_i =\frac{i}{2}\left(U_1\partial_i U_1^\dagger
+U_2\partial_i U_2^\dagger\right)\;\;+\;\;{\rm corrections}.
\la{sp1}\eeq
In the one-loop approximation it is sufficient to neglect corrections
to this value as they prove to be actually of the order of $g^2$, see
below.  This saddle-point value becomes exact if the kernel $K(x-y)$ is
a delta function considered as an example in ref. \cite{Zar1}. In ref.
\cite{KoKo1} more complicated technique has been involved to solve
the saddle-point equation than seems to be necessary for our goals
here.

It is amusing that `on the diagonal' where the gauge transformations
in $\Psi$ and the conjugate $\Psi^*$ coincide ($U_1=U_2$) the
saddle-point value $\bar A_i$ (that is where $|\Psi[A]|^2$ is mostly
concentrated) is a pure gauge potential. That is consistent with the
euclidean instanton-vacuum picture where the system spends much of the
`time' in a pure gauge state.

Actually, the introduction of two gauge transformations $U_1$ and
$U_2$ is abundant as one can always absorb one of them into a shift
of $A_i$ since the integration measure $DA_i$ is gauge invariant.
Renaming the integration variable $A_i^{U_1}\rightarrow A_i$
so that $A_i^{U_2}$ becomes $A_i^U$ where $U$ is the `relative'
gauge transformation, $U=U_1^\dagger U_2$, one can rewrite the norm
as

\beq
{\cal N}=\int\!DU\int\!DA\exp\left(-\frac{1}{g^2}Q[A]
-\frac{1}{g^2}Q[A^U]\right)
\la{norm2}\eeq
and the saddle-point field as

\beq
\bar A_i =\frac{i}{2}U\partial_i U^\dagger\;\;
+\;\;{\rm corrections}.
\la{sp2}\eeq
Generally speaking, this is not a pure gauge potential; it becomes
a pure gauge only when the relative gauge transformation $U$
does not differ much from unity.

Let us introduce the shifted field $A^\prime$ counted from the
saddle-point value \ur{sp2}, {\em i.e.} let us put
$A_i=(i/2)U\partial_iU^\dagger+A_i^\prime$ and expand $Q[A]+Q[A^U]$
in \eq{norm2} in powers of $A^\prime$.

The term quadratic in $A^\prime$ is

\[
\int\!\!\int\!d^3xd^3y\:K_{ij}(x-y)
\left\{2\Tr A_i^\prime(x)A_j^\prime(y)
+\Tr\left[U(y)U^\dagger(x)-1\right] A_i^\prime(x)
\left[U(x)U^\dagger(y)-1\right]A_j^\prime(y)\right.
\]
\beq \left.
+\Tr \left[U(y)U^\dagger(x)-1\right]A_i^\prime(x)A_j^\prime(y)
+\Tr A_i^\prime(x)\left[U(x)U^\dagger(y)-1\right]A_j^\prime(y)
\right\}.
\la{quad1}\eeq

The term linear in $A^\prime$ is

\[
\int\!\!\int\!d^3xd^3y\:K_{ij}(x-y)(-i)\Tr A_i^\prime(x)\left\{
\left[U(x)U^\dagger(y)-1\right]
\left[U(y)\partial_jU^\dagger(y)\right]
\left[U(y)U^\dagger(x)-1\right]\right.
\]
\beq\left.
+\left[U(x)U^\dagger(y)-1\right]
\left[U(y)\partial_jU^\dagger(y)\right]
+\left[U(y)\partial_jU^\dagger(y)\right]
\left[U(y)U^\dagger(x)-1\right]\right\}.
\la{lin1}\eeq

Finally, the term not containg at all $A^\prime$ is

\[
\int\!\!\int\!d^3xd^3y\:K_{ij}(x-y)\left\{
\frac{1}{2}\Tr \partial_i U^\dagger(x) \partial_jU(y)\right.
\]
\beq\left.
+\frac{1}{4}\Tr \partial_i U^\dagger(x) \partial_jU(y)
\left[U^\dagger(y)U(x)-1\right]
+\frac{1}{4}\Tr \partial_i U(y) \partial_jU^\dagger(x)
\left[U(x)U^\dagger(y)-1\right]\right\}.
\la{zero1}\eeq

We see that had $K_{ij}(x-y)$ been a $\delta$-function one could replace
$U(x)U^\dagger(y)$ by unity matrix, so that the linear term in
$A^\prime$ \ur{lin1} would vanish. In the same approximation one would
be left only with the first terms in \eqs{quad1}{zero1}. In fact this
approximation to $Q[A]+Q[A^U]$ is correct for any function
$K_{ij}(x-y)$ if we restrict ourselves to the one-loop approximation,
{\em i.e.} neglect $O(g^2)$ terms in calculating the norm ${\cal N}$.

To show that we first notice that if we, indeed, neglect the
linear term \ur{lin1} and leave only the first term in \eq{quad1}
and the first term in \eq{zero1}, the propagator of the $A_i^\prime$
field is of the order of $g^2$, and the propagator of the $U$ field is
also of the order of $g^2$. The last statement needs to be made more
precise as we, generally speaking, deal with a non-linear field: that
will be done below. What can immediately be seen is that the
fluctuations of the gradients of the $U$ field are of the order of
$g^2$, hence the characteristic square bracket $[U(x)U^\dagger(y)-1]$
is of that order, too, since it is non-zero only when one expands it
in the gradients of the $U$ field.

To check that we have correctly neglected all the terms we expand
$\exp(-Q[A]/g^2-Q[A^U]/g^2)$ in powers of the presumably
small terms. For example, let us consider the third and fourth terms
of \eq{quad1} expanded to the linear order. There is a factor $1/g^2$
coming with the expansion, a factor $g^2$ arising from the contraction
of the two $A^\prime$ fields and an extra factor $g^2$ owing to the
fluctuations of the square brackets, as explained above. We obtain thus
a $g^2$ correction, as compared to the main contribution to the norm
which is $O(1)$.

Let us consider the expansion in the linear term \ur{lin1}. Expanding
to the first order we get an identical zero (because it is odd in
$A^\prime$), so we have to consider the second term in the expansion of
the exponent.  We have $(1/g^2)^2$ from the expansion to the second
order, $g^2$ from the $A^\prime$ propagator and at least $g^4$ from
averaging over the $U$'s, that is again a $O(g^2)$ correction. The same
is true for the expansion of the last two terms in \eq{zero1}.

We conclude that if we are willing to neglect terms of the order of
$g^2$ ({\em i.e.} restricting ourselves to the one-loop
approximation) we can use the saddle-point field given by
\eqs{sp1}{sp2} and approximate the norm by a factorized expression:

\beq
{\cal N}\approx {\cal N}_A\cdot{\cal N}_U,
\la{norm3}\eeq
\beq
{\cal N}_A
=\!\int\!DA^\prime
\exp\left\{-\frac{2}{g^2}\int\!\!\int\!d^3x\:d^3y\:
K_{ij}(x-y)\:\Tr A_i^\prime(x)A_j^\prime(y)\right\},
\la{NA}\eeq
\beq
{\cal N}_U=\!\int\!DU\!
\exp\left\{-\frac{1}{2g^2}\int\!\!\int\!d^3x\:d^3y\:
K_{ij}(x-y)\:\Tr\partial_iU^\dagger(x)\partial_jU(y)\right\}.
\la{NU}\eeq

We see that the norm is factorized into independent integrals over
$A^\prime$ and over $U$. The former is a trivial gaussian integral
while the latter is non-trivial since the field $U$ is subject to
the unitarity constraint.

\section{Evaluation of the energy numerator
$\langle \Psi|{\cal H}|\Psi\rangle$}

We now proceed to evaluating the numerator of the vacuum energy
\ur{meanenergy}. We have

\beq
\langle \Psi|{\cal H}|\Psi\rangle
=\int\!DU\!\int\!DA\!
\int\!\!d^3x\frac{1}{2g^2}\left\{\frac{\delta Q[A]}{\delta A_i^a(x)}
\frac{\delta Q[A^U]}{\delta A_i^a(x)}+\left(B_i^a[A]\right)^2
\right\}\exp\left\{-\frac{1}{g^2}(Q[A]+Q[A^U])\right\} .
\la{num1}\eeq

The exponential factors are the same as for the norm integral, so we
have to analyze the prefactor which is new. Using \eq{deltaQ} we
get

\beq
\int\!d^3x\frac{\delta Q[A]}{\delta A_i^a(x)}
\frac{\delta Q[A^U]}{\delta A_i^a(x)}
=\int\!\!\int\!\!\int\!d^3xd^3yd^3zK_{ij}(x-y)K_{ik}(x-z)\:
2\:\Tr \left[U^\dagger(x)A_j(y)U(x)A_k^U(z)\right].
\la{deltaQdeltaQ}\eeq

We next expand this expression in the shifted field $A^\prime$
around the approximate saddle point substituting
$A_i=(i/2)U\partial_iU^\dagger+A_i^\prime$. To save space, we
write down the expansion of the last factor in \eq{deltaQdeltaQ}
only, the trace.

The term quadratic in $A^\prime$ is

\[
\Tr A_j^\prime(y)A_k^\prime(z)
+\Tr \left[U(z)U^\dagger(x)-1\right]A_j^\prime(y)
\left[U(x)U^\dagger(z)-1\right]A_k^\prime(z)
\]
\beq
+\Tr \left[U(z)U^\dagger(x)-1\right]A_j^\prime(y)A_k^\prime(z)
+\Tr A_j^\prime(y)\left[U(x)U^\dagger(z)-1\right]A_k^\prime(z).
\la{quad2}\eeq

The term linear in $A^\prime$ is

\beq
\frac{i}{2}\Tr A_k^\prime (z)
\left\{U(z)U^\dagger(x)U(y)\partial_jU^\dagger(y)
U(x)U^\dagger(z)-U(x)U^\dagger(y)U(y)\partial_jU^\dagger(y)
U(y)U^\dagger(x)\right\}.
\la{lin2}\eeq
At $y=x=z$ this term is zero implying that in the general case it
can be identically rewritten {\it \`a la} \eq{lin1} using expressions
of the type $\left[U(z)U^\dagger(x)-1\right]$. However, it leads to a
rather lengthy formula, and we do not present it here.

Finally, the term without the $A^\prime$ field is

\[
-\frac{1}{4}\left\{\Tr \partial_jU^\dagger(y)\partial_kU(z)
+\Tr \left[U^\dagger(x)U(y)-1\right]\partial_jU^\dagger(y)
\left[U(x)U^\dagger(z)-1\right]\partial_kU(z)\right.
\]
\beq\left.
+\Tr \left[U^\dagger(x)U(y)-1\right]\partial_jU^\dagger(y)\partial_kU(z)
+\Tr \partial_jU^\dagger(y)\left[U(x)U^\dagger(z)-1\right]
\partial_kU(z)\right\}.
\la{zero2}\eeq

We have now to integrate the sum of \eqss{quad2}{lin2}{zero2} over
the fields $A^\prime$ and $U$ with the exponential weight given by
\eqs{NA}{NU}. The leading $O(1)$ contributions come from the first
term in \eq{quad2} and from the first term in \eq{zero2}. \Eq{lin2}
can contribute only in combination with the linear term from the
exponent \ur{lin1} expanded to the first order: it gives a $g^2$
correction according to the power counting rules of the previous
section. By the same counting rules one can neglect all other terms
in \eqs{quad2}{zero2}.

A similar scenario takes place when we insert
$A_i=(i/2)U\partial_iU^\dagger+A_i^\prime$ into the magnetic
energy $B^2/2g^2$. The leading $O(1)$ contribution arises only from
the `abelian' piece made of the $A^\prime$ field, namely
from $\Tr(\partial_iA_j^\prime-\partial_jA_i^\prime)^2/2g^2$. All other
terms, including the Skyrme-type term made solely of the $U$ field,
give $O(g^2)$ corrections to the vacuum energy. Disregarding
systematically all such terms we finally obtain:

\[
\langle \Psi|{\cal H}|\Psi\rangle\approx
\int\!\!DU\!\!\int\!DA\exp\left\{-\frac{1}{g^2}\int\!\!\int\!
d^3x\:d^3y\:K_{ij}(x-y)\right.
\]
\[\left.
\cdot\left[2\Tr A_i(x)A_j(y)
+\frac{1}{2}\Tr\partial_iU^\dagger(x)\partial_jU(y)\right]\right\}
\]
\[
\times\frac{1}{g^2}\left\{
\int\!\!\int\!\!\int\!d^3xd^3yd^3zK_{ij}(x-y)K_{ik}(x-z)
\left[\Tr A_j(y)A_k(z)
-\frac{1}{4}\Tr \partial_jU^\dagger(y)\partial_kU(z)\right]\right.
\]
\beq\left.
+\frac{1}{2}\int\!d^3x\:\Tr(\partial_iA_j-\partial_jA_i)^2\right\}
\la{num2}\eeq
(we have deleted the primes in the notation of the $A$ field).

To get the variational estimate for the vacuum energy we need now
to divide \ur{num2} by the norm \ur{norm3}. Since the norm has a
factorized form the vacuum energy is a sum of two terms:

\beq
{\cal E}\geq \frac{\langle \Psi|{\cal H}|\Psi\rangle}
{\langle \Psi|\Psi\rangle}
=\left(\begin{array}{c}{\rm contribution}\\{\rm from\;\;}A\;\;
{\rm field} \end{array}\right)
+\left(\begin{array}{c}{\rm contribution}\\{\rm from\;\;}U\;\;
{\rm field} \end{array}\right)
\la{A+U}\eeq
where

\[
\left(\begin{array}{c}{\rm contribution}\\{\rm from\;\;}A\;\;
{\rm field} \end{array}\right)
=\frac{1}{g^2}\frac{\!\int\!DA\exp\left[...\right]}
{\!\int\!DA\exp\left[-\frac{2}{g^2}\int\!\!\int\!d^3x\:d^3y\:
K_{ij}(x-y)\:\Tr A_i(x)A_j(y)\right]}
\]
\beq
\cdot\left\{
\int\!\!\int\!\!\int\!d^3xd^3yd^3zK_{ij}(x-y)K_{ik}(x-z)
\Tr A_j(y)A_k(z)
+\frac{1}{2}\int\!d^3x\:\Tr(\partial_iA_j-\partial_jA_i)^2\right\},
\la{EnA}\eeq

\beq
\left(\begin{array}{c}{\rm contribution}\\{\rm from\;\;}U\;\;{\rm field}
\end{array}\right)
=-\frac{1}{4g^2}\frac{\int\!\!DU\exp\left[...\right]
\int\!\!\int\!\!\int\!dx\:dy\:dzK_{ij}(x-y)K_{ik}(x-z)
\Tr \partial_jU^\dagger(y)\partial_kU(z) }
{\int\!\!DU\exp\left[-\frac{1}{2g^2}
\int\!\!\int\!dx\:dy\:K_{ij}(x-y)
\Tr\partial_iU^\dagger(x)\partial_jU(y)\right] }.
\la{EU1}\eeq
Here $\exp[...]$ in both expressions mean the same as in the
corresponding denominators. The first term in \eq{EnA} is the
contribution of the $A_i$ field to the electric energy while
the second is the contribution to the magnetic energy. Notice that
in the one-loop approximation the $U$ field contributes only to the
electric energy.

The gaussian integrals over $A_i$ can be easily done in the momentum
space using the propagator

\beq
\langle A_i^a(p)A_j^b(-p)\rangle
=\frac{g^2}{2}\delta^{ab}K^{-1}_{ij}(p).
\la{Aprop}\eeq
Inversing the general expression \ur{Kp} for $K_{ij}(p)$ we have

\beq
K^{-1}_{ij}(p)=\frac{1}{a(p)}\left(\delta_{ij}-
\frac{p_ip_j}{p^2}\frac{b(p)}{a(p)+b(p)}\right).
\la{Kpinv}\eeq

We are now in a position to express the contribution of the
$A_i$ field to the electric and magnetic energy of the vacuum
through the trial functions $a(p)$ and $b(p)$.
Using \eqss{EnA}{Aprop}{Kpinv} we get

\[
\left(\begin{array}{c}{\rm magnetic}\\{\rm energy\;\;
density}\end{array}\right) =\frac{\langle B^2\rangle}{2g^2}
=\frac{N_c^2-1}{2g^2}\!\!\int\!\!\frac{d^3p}{(2\pi)^3}
(p^2\delta_{ij}-p_ip_j)\frac{g^2}{2}K^{-1}_{ij}(p)
\]
\beq
=\frac{N_c^2-1}{2}\!\!\int\!\!\frac{d^3p}{(2\pi)^3}\frac{p^2}{a(p)},
\la{BA}\eeq

\[
\left(\begin{array}{c}{\rm electric}\\{\rm energy\;\;
density}\end{array}\right) =\frac{\langle E^2\rangle}{2g^2}
=\frac{N_c^2-1}{2g^2}\!\!\int\!\!\frac{d^3p}{(2\pi)^3}
K_{ij}(p)K_{ik}(p)\frac{g^2}{2}K^{-1}_{jk}(p)\;+\;\ur{EU1}
\]
\beq
=\frac{N_c^2-1}{2}\!\!\int\!\!\frac{d^3p}{(2\pi)^3}\frac{3a(p)+b(p)}{2}
\;+\;\ur{EU1}.
%+\left(\begin{array}{c}{\rm contribution}\\{\rm from\;\;}U\;\;{\rm
%field} \end{array}\right).
\la{EA}\eeq

To write down the contribution of the $U$ field to the (electric)
energy density in a more compact way let us introduce new kernels,

\beq
L(x-y)=\partial_{xi}\partial_{yj}K_{ij}(x-y),\;\;\;\;\;
L(p)=p_ip_jK_{ij}(p)=p^2\left[a(p)+b(p)\right]
\la{L}\eeq
and

\beq
S(y-z)=\int\!d^3x\partial_{yj}K_{ij}(x-y)\:\partial_{zk}K_{ik}(x-z),
\;\;\;\;S(p)=p^2\left[a(p)+b(p)\right]^2=\frac{L^2(p)}{p^2}.
\la{S}\eeq
\Eq{EU1} can be compactly rewritten through the variational derivative
of the norm,

\beq
\left(\begin{array}{c}{\rm contribution}\\{\rm from\;\;}U\;\;{\rm field}
\end{array}\right)
=\frac{1}{2}\!\!\int\!\!\frac{d^3p}{(2\pi)^3}S(p)
\frac{\delta\log{\cal N}_U}{\delta L(p)}\frac{(2\pi)^3}{V},
\la{EU2}\eeq
where

\beq
{\cal N}_U=\int\!DU\exp\left[-\frac{1}{2g^2}
\!\!\int\!\!\frac{d^3p}{(2\pi)^3}L(p)\Tr U^\dagger(p)U(-p)\right].
\la{normU}\eeq

In the next two sections we shall study the contribution of the $U$
field to the electric energy density in two regimes: perturbative,
which will serve as a pedagogical example, and non-perturbative.

\section{Perturbative regime}

In the perturbative regime one assumes that the integration over the
unitary field $U$ in \eqs{NU}{normU} is concentrated in the vicinity
of unity matrices. Hence we can expand $U$ as

\beq
U=\exp(i\phi^at^a)\approx 1+i\phi^at^a,\;\;\;\;
U^\dagger\approx 1-i\phi^at^a,\;\;\;\;\;DU=D\phi^a.
\la{Uexpan}\eeq

In this approximation the action for the $N_c^2-1$ fields $\phi^a$
becomes quadratic, leading to the propagator

\beq
\langle \phi^a(p)\phi^b(-p)\rangle
=\frac{2g^2\delta^{ab}}{L(p)}
=\frac{2g^2\delta^{ab}}{p^2[a(p)+b(p)]}.
\la{phiprop}\eeq

The addition to the electric energy density arising from the $U$ field
is thus

\[
\left(\begin{array}{c}{\rm perturbative\;\;contribution}\\
{\rm from\;\;the\;\;}U\;\;{\rm field} \end{array}\right)
=-\frac{N_c^2-1}{8g^2}\!\!\int\!\!\frac{d^3p}{(2\pi)^3}
S(p)\frac{2g^2}{L(p)}
\]
\beq
=-\frac{N_c^2-1}{2}\!\!\int\!\!\frac{d^3p}{(2\pi)^3}
\frac{L(p)}{2p^2}
=-\frac{N_c^2-1}{2}\!\!\int\!\!\frac{d^3p}{(2\pi)^3}
\frac{a(p)+b(p)}{2}.
\la{Ephipert}\eeq
Naturally, this result comes also from the general formula \ur{EU2}.

Adding this result to \eq{EA} we obtain

\beq
\left(\begin{array}{c}{\rm electric}\\{\rm energy\;\;
density}\end{array}\right) =\frac{\langle E^2\rangle}{2g^2}
=\frac{N_c^2-1}{2}\!\!\int\!\!\frac{d^3p}{(2\pi)^3}a(p).
\la{Efullpert}\eeq
Notice the cancellation of the longitudinal piece associated with
the trial function $b(p)$. This is as it should be in the perturbative
regime.

We now add up the electric \ur{Efullpert} and the magnetic \ur{BA}
parts of the vacuum energy and get the variational estimate
for the vacuum energy: \footnote{Actually, \eq{Enpert} can be
found in ref. \cite{KoKo1}.}

\beq
\frac{{\cal E}}{V}
\geq \frac{N_c^2-1}{2}\!\!\int\!\!\frac{d^3p}{(2\pi)^3}
\left[a(p)+\frac{p^2}{a(p)}\right].
\la{Enpert}\eeq
This expression should be varied in respect to the trial function
$a(p)$. The r.h.s. of \eq{Enpert} has the minimum at

\beq
a(p)=|p|,\;\;\;\;\;b(p)=\;\;{\rm arbitrary},
\la{apert}\eeq
the vacuum energy at the minimum being

\beq
\frac{{\cal E}}{V}
= 2(N_c^2-1)\!\!\int\!\!\frac{d^3p}{(2\pi)^3}\frac{|p|}{2}.
\la{Enminpert}\eeq
This is, of course, the correct energy of the zero-point oscillations
of $N_c^2-1$ free gluons with two physical polarizations, each mode
with momentum $p$ carrying the energy $|p|$. Note that at the minimum
the virial theorem for harmonic oscillators is satisfied, since
$\langle E^2\rangle=\langle B^2\rangle$.

These nice results do not come unexpected since the gaussian wave
functional used here is exact for the free fields. We have only shown
that its exact form given by \eq{apert} can be found from the
variational principle.

\section{Key point: non-zero v.e.v. of the Lagrange multiplier}

Let us consider the $SU(2)$ gauge group. The unitary matrix $U$
in \eq{EU1} can be parametrized by quaternions,

\beq
U=u_\alpha\sigma^-_\alpha,\;\;\;\;\;
U^\dagger=u_\alpha\sigma^+_\alpha,\;\;\;\;\;
\sigma^\pm_\alpha=(\pm i\tau, 1),\;\;\;\;\;u_\alpha^2=1.
\la{quatern}\eeq
The last condition imposed on four real fields $u_\alpha$ ensures
that $U$ is a matrix from $SU(2)$.

The functional integral over $SU(2)$ matrices $U$ can be written
as

\beq
\int\!DU = \int\!Du_\alpha\prod_x\delta\left(u_\alpha^2(x)-1\right)
=\int\!Du_\alpha\int\!D\lambda\exp\left[-\frac{1}{g^2}\int\!d^3x
(\lambda u_\alpha^2-\lambda)\right]
\la{Lagr}\eeq
where we have introduced an auxiliary integration over `Lagrange
multiplier' field $\lambda(x)$ to ensure that the unitarity condition
is satisfied at all space points. The factor $1/g^2$ is inserted in
\eq{Lagr} for future convenience. The integration measure \ur{Lagr}
is invariant under left and right shifts along the group, therefore
it belongs to the class of the Haar measures.

The dependence of the energy numerator
$\langle \Psi|{\cal H}|\Psi\rangle$ and of the denominator
$\langle \Psi|\Psi\rangle$ on the $U$ field enters through the
expression (see \eq{EU1})

\beq
\Tr \partial_iU^\dagger(x)\partial_jU(y)
=2\partial_iu_\alpha(x)\partial_ju_\alpha(y).
\la{expo1}\eeq
Therefore, the norm \ur{NU} or \ur{normU} reads

\beq
{\cal N}_U=
\int\!D\lambda\exp\left(\frac{1}{g^2}\int\!d^3x\:\lambda\right)\!\!
\int\!\!Du_\alpha\exp\left\{-\frac{1}{g^2}\left[\int\!\!
\int\!d^3xd^3y L(x-y)u_\alpha(x)u_\alpha(y)
+\int\!d^3x\lambda u_\alpha^2\right]\right\}
\la{normU2}\eeq
where we have used the new kernel $L$ as defined by \eq{L}.
The inner integral is a gaussian integral over four real fields
$u_\alpha$ with the Lagrange multiplier $\lambda$ playing the role of
the mass term.

We have learned from the previous section that $a(p)=|p|$ at large
momenta. The behaviour of the trial function $b(p)$ is not defined
by the leading-order perturbation theory, however one can reasonably
assume that it can be also $\sim |p|$. Therefore, the asymptotics of
the `kinetic energy' term in \eq{normU2} is

\beq
L(p)\rightarrow c|p|^3,\;\;\;\;\;c=1+b,
\la{Lasym}\eeq
where $b$ is the (unknown) numerical coefficient in $b(p)\rightarrow
b|p|$ at large momenta.

Although the number of fields is only four ($n=4$) we shall treat it
as a formal parameter. The gaussian integral over $n$ real fields
$u_\alpha$ in \eq{normU2} can be formally written as

\beq
\int\!Du_\alpha\exp\left\{...\right\}
=\exp\left[-\frac{n}{2}\Sp\log(L+\lambda)\right]
\la{gauss1}\eeq
where $\Sp$ stands for the functional trace. Let us divide the Lagrange
multiplier field $\lambda(x)$ into a point-independent part $\mu^3$
and a space-variable part $\lambda^\prime(x)$:

\beq
\lambda(x)=\mu^3+\lambda^\prime(x),\;\;\;\;\;
\int\!d^3x\:\lambda^\prime(x)=0.
\la{lambdaprime}\eeq
The functional trace in \ur{gauss1} can be rewritten as

\beq
\Sp\log(L+\lambda)=V\!\int\!\frac{d^3p}{(2\pi)^3}
\log\left[L(p)+\mu^3\right]
-\frac{1}{2}\!\int\!\frac{d^3q}{(2\pi)^3}\lambda^\prime(q)\Pi(q)
\lambda^\prime(-q)
+O(\lambda^{\prime 3})
\la{Spur}\eeq
with $\Pi(q)$ being a loop made of the $u_\alpha$ propagators:

\beq
\Pi(q)=\!\int\!\frac{d^3p}{(2\pi)^3}
\left[\frac{1}{[L(p+q/2)+\mu^3][L(p-q/2)+\mu^3]}
-\frac{1}{[L(p)+\mu^3][L(p)+\mu^3]}\right].
\la{Piq}\eeq

\Eq{normU2} defines a peculiar $O(n)$ sigma model in three
dimensions at $n=4$.  Were $n$ a large parameter the fluctuations of
the $\lambda^\prime$ field would be suppressed as $1/n$. In that case
one could neglect, to the first approximation in $1/n$, the second and
higher terms in \eq{Spur}. Though $n=4$ cannot be considered as
particularly `large' we shall nevertheless leave out the quadratic and
higher-order terms in $\lambda^\prime$. In principle, these terms can
be taken into account systematically in the $1/n$ expansion similarly
to the way it is done in two-dimensional sigma models
\cite{Aref,RimWeis,Ter}.

The dependence on the constant part of the Lagrange multiplier
$\mu^3$ is given by (see \eqs{normU2}{Spur})

\beq
V\cdot\left\{\frac{\mu^3}{g^2}
-\frac{n(=4)}{2}\!\int\!\frac{d^3p}{(2\pi)^3}
\log\left[L(p)+\mu^3\right]\right\}.
\la{mudep}\eeq
This expression has an extremum in $\mu$ determined by the `gap'
equation

\beq
\frac{1}{g^2}-2\!\int\!\frac{d^3p}{(2\pi)^3}
\frac{1}{L(p)+\mu^3}=0,
\la{extrem}\eeq
the integral here being logarithmically divergent since $L(p)
\approx c|p|^3$ at large $|p|$ (see \eq{Lasym}). Introducing an
ultraviolet cutoff $M$ we get from \ur{extrem} the vacuum expectation
value of the Lagrange multiplier $\bar\lambda=\mu^3$ where

\beq
\mu={\rm const}\:M\exp\left[-\frac{8\pi^2}{g^2\:2\:(4/c)}\right].
\la{reninv}\eeq
Comparing it with the asymptotic-freedom law \ur{Lambda} we see that
in order to get the correct one-loop Gell-Mann--Low coefficient
(at $N_c=2$) we need to put

\beq
\frac{4}{c}=\frac{11}{3}\;\;\;\;\;{\rm or}\;\;\;\;\;
c=1+\left.\frac{b(p)}{|p|}\right|_{p\rightarrow\infty}
=\frac{12}{11}.
\la{12/11}\eeq

We notice that the first Gell-Mann--Low coefficient has been recently
calculated in ref. \cite{BrownKo} in the Kogan--Kovner ansatz
using a different method, and it has been found to be $12/3$
instead of $11/3$. This is exactly what we get from \eqs{reninv}{12/11}
if we deliberately put (as it has been done in \cite{BrownKo})
$b(p)=0$. However we emphasize again that there are no reasons to bind
oneself with such restriction. The actual value of the trial function
$b(p)$ at large $|p|$ should come from minimizing the
{\em second}-order $O(M^4g^2)$ term in \eq{Eexpansion}. We have
systematically neglected such terms above, however they can be
systematically collected as well. Those terms contain, in particular,
the non-abelian commutator terms for the magnetic energy, involving
`transverse' gluons. Therefore, it can well happen that the `best'
$b(p)\rightarrow |p|/11$, as it should be for getting the correct
Gell-Mann--Low coefficient.

\Eq{reninv} demonstrates how transmutation of dimensions occurs
in the Schr\"odinger approach. We see that it is related to the
non-zero vacuum expectation value of the Lagrange multiplier field
needed to get rid of the unitarity restriction for gauge
transformations. This phenomenon is similar to what happens in the
$O(n)$ sigma models in two dimensions. Similarly to those models one
needs, strictly speaking, the large $n$ parameter to justify the use
of the saddle-point method. However, one can hope that $n=4$ is
sufficiently `large' so that the conclusion that the Lagrange
multiplier field $\lambda$ does get a non-zero v.e.v. is not altered
by fluctuations of the $\lambda^\prime$ field about the saddle-point.
These fluctuations can be anyhow treated systematically in a formal
$1/n$ expansion.

It is instructive to separate the perturbative and nonperturbative
contributions to the vacuum energy by adding and subtracting the
perturbative contribution \ur{Ephipert}. We have

\[
\frac{{\cal E}}{V}\geq\frac{{\cal E}^{pert}}{V}+
\frac{{\cal E}^{nonpert}}{V}
\]
\beq
=2(N_c^2-1)\!\!\int\!\!\frac{d^3p}{(2\pi)^3}\frac{|p|}{2}\;\;\;
+\;\;\;\int\!\!\frac{d^3p}{(2\pi)^3}\left[(N_c^2-1)\frac{L(p)}{2p^2}
-\frac{L^2(p)}{2p^2}\frac{\delta\log{\cal N}_U}{\delta L(p)}
\frac{(2\pi)^3}{V}\right].
\la{Ep+Enp}\eeq

${\cal E}^{nonpert}$ is a functional of the kernel $L(x-y)$ only
(see \eq{L}), and the variational principle comes to minimizing
this functional. To perform that one needs to evaluate the norm
${\cal N}_U$ \ur{normU} for arbitrary $L(x-y)$ and then mimimize
the r.h.s. of \eq{Ep+Enp} in $L$. After the transmutation of dimensions
takes place, the bare coupling $g^2$ disappears from all expressions,
being replaced by a mass parameter $\mu\sim\Lambda$, see
\eq{reninv}. In momentum space $L(p,\mu)$ should asymptotically go
to $(12/11)|p|^3$ at $|p|\gg\mu$, however this behaviour cannot follow
from the minimization of the nonperturbative vacuum energy but
should be rather imposed `by hands' as presumably following
from the minimization of the second-order perturbative energy.
Apart from this boundary condition, the shape of $L(p,\mu)$ is to be
extracted from the variation of ${\cal E}^{nonpert}$ which has to be
convergent at large momenta. The whole procedure seems to be feasible
in the framework of the $1/n$ expansion.

In the leading order of the $1/n$ expansion one freezes the Lagrange
multiplier $\lambda$ at its saddle-point value $\bar\lambda=\mu^3$
and neglects the fluctuations of the $\lambda^\prime$ field. The
propagator of the $u_\alpha$ field becomes

\beq
\langle
u_\alpha(p)\:u_\beta(-p)\rangle=\frac{g^2}{2}
\delta_{\alpha\beta}\frac{1}{L(p)+\mu^3}.
\la{uprop}\eeq

Concluding this section we would like to make the following remark.
In refs. \cite{KoKo1,Zar1} another method of circumventing the
unitarity condition has been suggested. In these refs. integration over
unitary matrices $U$ has been replaced by

\beq
\int\!DU=\int\!DU^\alpha_\beta DU^{*\alpha}_\beta\:
\delta\left([U^\dagger U]^\alpha_\beta-\delta^\alpha_\beta\right)
=\int\!D\sigma_\beta^\alpha\!\int\!DU^\alpha_\beta DU^{*\alpha}_\beta\:
\exp\left\{-\int\!dx\:\Tr\sigma (U^\dagger U-1)
\right\}.
\la{drugojLagr}\eeq
Here one has $N_c^2$ complex fields $U,U^\dagger$ with $2N_c^2$ real
degreees of freedom (neglecting the difference between the $SU(N_c)$
and $U(N_c)$ groups), with $N_c^2$ restrictions taken care of by
integrating over $N_c^2$ Lagrange multiplier fields
$\sigma=\sigma^\dagger$. Taking $N_c\rightarrow\infty$ does not help
to justify any saddle point in $\sigma$, as it is assumed in
refs. \cite{KoKo1,Zar1}. Indeed, the effective action in $\sigma$
resulting from integrating over the unconstrained fields $U,U^\dagger$
depends not only on $N_c$ eigenvalues of $\sigma^\alpha_\beta$ varying
slowly but also on $N_c^2-N_c=O(N_c^2)$ `angle' variables of
$\sigma^\alpha_\beta$ which fluctuate violently. Meanwhile, the
fluctuations of the `angle' part of $\sigma^\alpha_\beta$ influence
the effective action for the eigenvalues of the $\sigma$ field, and
taking $N_c\rightarrow\infty$ does not quench the `angle' variables
as their number is $O(N_c^2)$. In short, the method does not work.

In order to get use of the large $N_c$ parameter one has to
parametrize the $SU(N_c)$ group by introducing $O(N_c^2)$ `flat'
variables subject to $O(N_c)$ constraints. In the case of $N_c=2$
it has been explicitly performed above. For larger $N_c$ we are,
unfortunately, unaware of such a parametrization.

Finally, it should be mentioned that the second derivative of the
effective potential \ur{mudep} in $\lambda$ is positive-definite,
which may seem to invalidate the use of the saddle-point method. In
fact, it is exactly the needed sign of the curvature, since integration
over the Lagrange multiplier field $\lambda$ in \eq{normU2} actually
goes along the {\em imaginary} axis. The integration path climbs to
the maximal value of the integrand (which happens to be exactly on the
real axis!) and then descends, as it should be in the saddle-point
integration.

\section{A queer way to get the linear potential}

Let us now consider two static colour charges in the fundamental
representation on top of the vacuum state. We shall use the trial wave
functional suggested by Zarembo \cite{Zar1}. Apart from being a
functional of the $A_i$ field, it is a function of where we put the
probe quark ($z_1$) and antiquark ($z_2$); it is also a colour matrix:

\beq
\Psi^\alpha_\beta[A;\:z_1,z_2]=\int\!DU\:U^\alpha_\gamma(z_1)
U^{\dagger \gamma}_\beta(z_2)
\int\!DA\exp\left(-\frac{Q[A^U]}{g^2}\right).
\la{wfistoch}\eeq
One can immediatelly check that under gauge transformations
the wave functional \ur{wfistoch} satisfies the needed condition
\ur{gauge-cov}. As correctly stressed in ref. \cite{Zar1} the static
potential between probe quarks is anyhow a property of the vacuum state
and it does not matter much what precisely trial functional do we use
to `measure' it, provided it satisfies the Gauss condition
\ur{gauge-cov}. \Eq{wfistoch} is a simple and natural choice.

By definition, the potential between static sources is the
hamiltonian averaged with the wave functional \ur{wfistoch} minus
the vacuum energy:

\beq
V(z_1-z_2)=\frac{\langle \Psi(z_1,z_2)|{\cal H}|\Psi(z_1,z_2)\rangle}
{\langle \Psi(z_1,z_2)|\Psi(z_1,z_2)\rangle}
-\frac{\langle \Psi_0|{\cal H}|\Psi_0\rangle}
{\langle \Psi_0|\Psi_0\rangle}.
\la{Vdef}\eeq
The norm of the $\Psi(z_1,z_2)$ state is \cite{Zar1}

\beq
{\cal N}(z_1-z_2)=\langle \Psi(z_1,z_2)|\Psi(z_1,z_2)\rangle
=\int\!DU\!\int\!DA\:\Tr \left[ U(z_1)U^\dagger (z_2)\right]\:
\exp\left(-\frac{Q[A]}{g^2}-\frac{Q[A^U]}{g^2}\right)
\la{normz}\eeq
where we have again, as in the vacuum case, introduced the relative
gauge transformation $U=U_1^\dagger U_2$. Performing the same one-loop
approximations as in sections 5 and 6, it is easy to see that the
contributions from the $A_i$ field get cancelled in \eq{Vdef},
and we are left with the correlation functions of the sigma model.
Let us introduce a short-hand notation for these correlation functions:

\beq
\langle\!\langle {\cal O} \rangle\!\rangle
=\frac{\int\!DU\exp\left(-\frac{1}{2g^2}\!\int\!\!\int\!dx\:dy\:
L(x-y)\:\Tr U^\dagger(x)U(y)\right){\cal O}}
{\int\!DU\exp\left(-\frac{1}{2g^2}\!\int\!\!\int\!dx\:dy\:
L(x-y)\:\Tr U^\dagger(x)U(y)\right)}.
\la{llrr}\eeq

The static potential can be written as

\[
V(z_1-z_2)=-\frac{4}{g^2}\left[\frac{\int\!\!\int\!dx\:dy\:S(x-y)\:
\laa\Tr U(z_1)U^\dagger(z_2)\Tr U^\dagger(x)U(y)\raa}
{\laa \Tr U(z_1)U^\dagger (z_2) \raa}\right.
\]
\beq
\left.-\;\frac{\int\!\!\int\!dx\:dy\:S(x-y)\:
\laa\Tr U^\dagger (x)U(y)\raa}{\laa 1 \raa}\right].
\la{V1}\eeq
In the leading order of the $1/n$ expansion the integrals in
\eqs{llrr}{V1} become gaussian with the propagator given by \eq{uprop}.
There are three possible contractions of the $u_\alpha$ fields in the
numerator of \eq{V1}: one of them gets cancelled by vacuum subtraction
in \ur{V1}; the other two are equivalent. Using the $u_\alpha$
propagator \ur{uprop} we obtain

\[
V(r)=\frac{E(r)}{N(r)},\;\;\;\;
N(r)=2\!\int\!\frac{d^3p}{(2\pi)^3}\:e^{i(p\cdot r)}
\frac{1}{L(p)+\mu^3}
=\frac{2}{2\pi^2}\!\int_0^\infty dp\:\frac{\sin(pr)}{p}
\frac{1}{L(p)+\mu^3}.
\]
\beq
E(r)=-\int\!\frac{d^3p}{(2\pi)^3}\:e^{i(p\cdot r)}
\frac{L^2(p)}{p^2[L(p)+\mu^3]^2}
=-\frac{1}{2\pi^2}\!\int_0^\infty dp\:p\:\sin(pr)
\left(\frac{L(p)}{L(p)+\mu^3}\right)^2,
\la{V2}\eeq

Let us first investigate the short-distance behaviour of $V(r)$.
Small $r$ correspond to large $p$ where $L(p)=(12/11)|p|^3$ according
to \eq{12/11}. We have at small $r$

\beq
E(r)\rightarrow -\frac{1}{4\pi r},\;\;\;\;\;\;
N(r)\rightarrow \frac{11}{12}\;\frac{1}{\pi^2}\;\ln\frac{1}{r},
\;\;\;\;\;\;V(r)\rightarrow -\frac{1}{4\pi r}\;
\frac{8\pi^2}{\frac{22}{3}\ln\frac{1}{r}}.
\la{smallr}\eeq
This is almost the correct Coulomb potential modified by the running of
the gauge coupling constant. The only difference with the true
asymptotics at small $r$ is the absence of the $(N_c^2-1)/(2N_c)=3/4$
colour factor in \eq{smallr}. The difference $1-3/4=1/4$ is due to our
use of the leading order of the $1/n$ expansion at $n=4$;
simultaneously it gives the idea of the numerical error.

The behaviour of $V(r)$ at large separations depends crucially on the
{\em analytic properties} of $L(p)$. The asymptotics of the Fourier
transforms \ur{V2} are defined by the singularities of the integrands
in $p^2$. If $L(p,\mu)+\mu^3$ has a singularity at $p^2=0$ (for
example, is odd in $|p|$) the large-$r$ asymptotics of both $E(r)$
and $N(r)$ are determined by the integration origin at $|p|=0$:
both quantities exhibit a power decay at large $r$. Examples
we have considered all show that their ratio $V(r)$ is also decreasing
as a power of $r$. This is not what we want for confinement.

If $L(p,\mu)+\mu^3$ is even in $|p|$ one can expand the integration
range in $|p|$ in \eq{V2} from $(0,\infty)$ to $(-\infty,\infty)$.
The asymptotics will be then determined by the positions of the
singularities of the integrands in the $|p|$ plane. If these
singularities have a real part both $E(r)$ and $N(r)$ and consequently
$V(r)$ are oscillating. Probably, this is also unacceptable.

If $L(p,\mu)+\mu^3$ has zeros on the imaginary axis of $|p|$, say
at $|p|=\pm i\kappa$, $E(r)$ and $N(r)$ decay exponentially
as $\sim\exp(-\kappa r)$. The key point is that if the integrand
of $N(r)$ has a single pole at $|p|=\pm i \kappa$, the integrand of
$E(r)$ has a {\em double} pole. Taking the residue of the double pole
one can differentiate $\exp(i|p|r)$ and get an extra factor of $r$
in $E(r)$ as compared to $N(r)$. The same mathematical mechanism
works if $L(p,\mu)+\mu^3$ has a cut at some imaginary value of $|p|$.
This is the way one obtains an infinitely rising linear potential.

To give an example we choose

\beq
L(p,\mu)+\mu^3=\frac{12}{11}\left[p^2+\left(\frac{11}{12}\right)
^{\frac{2}{3}}\mu^2\right]^{\frac{3}{2}}
\la{example}\eeq
satisfying the above considerations and also the asymptotics
$(12/11)|p|^3$. The coefficient inside the square brackets has been
chosen so that $b(p)$ is not singular at $p=0$. The corresponding
potential $V(r)$ is plotted in Fig.1.

%%%%%%%%%%
% FIGURE %
%%%%%%%%%%
\begin{figure}
%\vspace*{}
\centerline{\epsfxsize10.0cm\epsffile{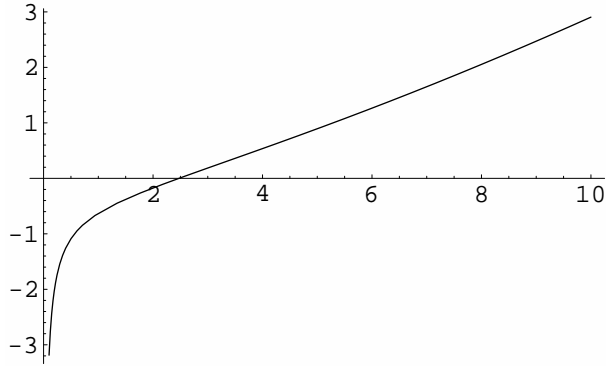}}
%\vspace*{}
\caption[]{Static potential following from \eq{example}.
The axes are in units of $\mu$ obtained from the
transmutation of dimensions. The standard string tension
$\sigma\approx (440\;MeV)^2$ corresponds to $\mu\approx 600\;MeV$.}
\end{figure}

Both the Coulomb part \ur{smallr} and the linear part are clearly
seen. In mathematical terms, both $E(r)$ and $N(r)$ decay exponentially
as $\sim \exp(-{\rm const}\:\mu\:r)$ but $N(r)$ is {\em additionally
suppressed} as $1/r$. Being inversed, it gives a linear rising
potential.

Let us put it in a more general way: In the original formula for
the potential \ur{Vdef} both the numerator and the denominator have
to decay exponentially if there is a mass gap in the theory. However
the denominator, that is the norm of the state with colour sources
$\langle \Psi(z_1,z_2)|\Psi(z_1,z_2)\rangle$, has to fall off as
a factor $1/r$ faster than the average of the hamiltonian,
$\langle \Psi(z_1,z_2)|{\cal H}|\Psi(z_1,z_2)\rangle$. Of course, one
can always artificially normalize the wave functional to unity
but that does not seem to be a natural thing to do.

We have called this way of getting confinement `queer'
because it is very different from the usual Euclidean standpoint where
one has to demonstrate a tremendously small $\exp(-{\rm Area})$
behaviour of the Wilson loop or a similar behaviour of the Polyakov's
loops correlator.  This is very difficult both theoretically and in
practice. However in the Schr\"odinger approach getting the linear
potential is not at all queer but rather quite natural.

\section{Adjoint sources}

Let us first of all introduce the gauge transformation in the adjoint
representation:

\beq
O^{ab}(U)=2\Tr \left[ U^\dagger t^a U t^b \right],\;\;\;\;\;\;
O^{ac}O^{bc}=O^{ca}O^{cb}=\delta^{ab}.
\la{adj1}\eeq

We take the wave functional for a state with two adjoint sources
sitting at points $z_{1,2}$ in the form similar to \eq{wfistoch}:

\beq
\Psi^{ab}[A;\:z_1,z_2]=\int\!DU\:O^{ac}(z_1)O^{bc}(z_2)
\int\!DA\exp\left(-\frac{Q[A^U]}{g^2}\right).
\la{wfadj}\eeq
It satisfies the needed gauge transformation law,

\beq
\Psi^{ab}[A^S;\:z_1,z_2]= \left[O^{-1}(S(z_1))\right]^{aa^\prime}
\Psi^{a^\prime b^\prime}[A;\:z_1,z_2]
\left[O(S(z_2))\right]^{b^\prime b}.
\la{adjtransf}\eeq

Proceeding in the same fashion as in the previous section we arrive to
the following expression for the potential for static adjoint charges:

\[
V^{adj}(z_1-z_2)=-\frac{4}{g^2}\left[\frac{\int\!\!\int\!dx\:dy\:S(x-y)\:
\laa\left[\Tr U^\dagger(z_1)U(z_2)\Tr U(z_1)U^\dagger(z_2)-1\right]
\Tr U^\dagger(x)U(y)\raa}
{\laa \Tr U^\dagger(z_1)U(z_2)\Tr U(z_1)U^\dagger(z_2)-1\raa}\right.
\]
\beq
\left.-\;\frac{\int\!\!\int\!dx\:dy\:S(x-y)\:
\laa\Tr U^\dagger (x)U(y)\raa}{\laa 1 \raa}\right].
\la{V1adj}\eeq

Passing from integration over unitary matrices $U$ to the flat
variables $u_\alpha$ and the Lagrange multiplier $\lambda$ one observes
that there are much more possible contractions of $u_\alpha$ fields
than in the case of fundamental sources. Not going into details here we
mention that thanks to the `gap' equation \ur{extrem} contributions
potentially possessing the confining behaviour get cancelled. There is
no linear potential between static adjoint sources even if there is
such for the fundamental charges.

\section{What is to be done next}

We believe that the variational approach may become a powerful
alternative to the Euclidean study of gluodynamics. To put it on a more
solid basis one needs:

\begin{itemize}
\item To compute within the variational ansatz given by
\eqss{KoKo}{Q}{Kp} the first perturbative correction to the vacuum
energy, which is of the order of $g^2M^4$. Only the asymptotics of the
trial functions $a(p),\;b(p)$ at large $|p|$ enter this calculation.
The function $a(p)=|p|$ is determined from the first-order calculation
(see above), the function $b(p)$ is not. Computing the $O(g^2)$
correction will fix this function. There is a good chance that the
best $b(p)$ will turn out to be $|p|/11$ as required by asymptotic
freedom, \eq{12/11}.

\item To compute $1/n$ corrections to the saddle-point approximation
to the sigma model \ur{normU2}. It will enable one to learn, by
minimizing the r.h.s. of \eq{Ep+Enp}, the nonperturbative propagator
$L(p,\mu)+\mu^3$.

\end{itemize}

Both computations seem to be feasible. There are also obvious
applications of the variational approach:

\begin{itemize}
\item The glueball masses and wave functions
\item The $\theta$ dependence of the vacuum energy, or the topological
susceptibility of the vacuum
\item The area behaviour of the spatial Wilson loop (?)
\end{itemize}

\section{Conclusions}

The gaussian variational ansatz for the vacuum wave functional is, on
one hand, sufficiently simple to allow its study by conventional
field-theoretic methods but on the other hand, it is probably sufficiently
`rich' to incorporate many desired features of the Yang--Mills theory.

We have shown that by including the longitudinal structure in the
gaussian kernel one can, in principle, get the correct one-loop
$\beta$-function of the Yang--Mills theory within the variational
approach.

The famous transmutation of dimensions arises very natural in the
Schr\"odinger picture. It results from a phenomenon similar to what
is well known in $O(n)$ sigma models in two dimensions, namely from
the appearance of a nonzero vacuum expectation value of the Lagrange
multiplier field needed to get rid of the unitarity constraint on the
gauge transformations. We have performed it explicitly for the $SU(2)$
gauge group; unfortunately we are unable to generalize the method to
higher $N_c$.

The appearance of the mass gap through the transmutation of dimensions
is good news, however it is not sufficient by itself to get a linear
potential between source quarks. The confinement requirement can be
formulated as a simple analyticity property of the propagator in the
sigma model, once the Lagrange multiplier gets a nonzero v.e.v. This
propagator can be systematically studied within the $1/n$ expansion,
again similar to what has been developed in $d=2$ models.

The variational approach allows certain freedom in action. Instead
of working hard trying to establish the best parameters / functions
within a variational ansatz, one can choose them to be what one likes,
for example, taking the propagator in the form of \eq{example} leading
to a beautiful static potential shown in Fig.1. After all, the
variational principle says that one gets an upper bound for the energy
for any trial function. Using, for example, \eq{example} it is
interesting to check if it leads to reasonable glueball states,
reasonable topological susceptibility, and so on.

The most challenging theoretical question is whether the
variational wave functional gives rise to the area law for the spatial
Wilson loop. To check it, the representation of the Wilson loop via
the nonabelian Stokes theorem of ref. \cite{DPStokes} would be probably
helpful.

\vskip 1true cm
{\bf Acknowledgements}
\vskip .8true cm
I am grateful to V.Yu.Petrov for useful discussions and to
Ya.I.Azimov and R.M.Ryndin for organizing the PNPI Winter School
where this lecture has been presented.

\end{document}